\documentclass[12pt]{article}
\usepackage{amsmath,amssymb}
\usepackage{cite}
\numberwithin{equation}{section}

\newcommand{\Rb}{\mathbb{R}}
\newcommand{\Zb}{\mathbb{Z}}
\newcommand{\Ncal}{{\cal N}}
\newcommand{\del}{\partial}
\newcommand{\Acal}{{\cal A}}
\newcommand{\nn}{\nonumber}
\newcommand{\taub}{\bar{\tau}}
\newcommand{\im}{{\rm Im}\,}
\newcommand{\Xt}{\tilde{X}}
\newcommand{\Mt}{\tilde{M}}
\newcommand{\Phis}{{*\Phi}}
\newcommand{\As}{*A}
\newcommand{\Hs}{*H}

\newcommand{\G}[1]{\mbox{$G_{#1}$}}
\newcommand{\Spin}[1]{{\rm Spin}(#1)}
\newcommand{\SO}[1]{{\rm SO}(#1)}
\newcommand{\SU}[1]{{\rm SU}(#1)}
\newcommand{\psil}{\psi_{\phi}}
\newcommand{\kh}{\hat{k}}
\newcommand{\Jh}{J}
\newcommand{\Jt}{\tilde{J}}
\newcommand{\Jf}{J_{\rm f}}
\renewcommand{\k}{k}
\newcommand{\hc}{{h^{\vee}}}
\newcommand{\Hf}{\hat{H}_{\rm f}}
\newcommand{\isi}{{\rm Ising}}
\newcommand{\tri}{{\rm Tri}}
\newcommand{\Bcal}{{\cal B}}
\newcommand{\h}[1]{#1}
\newcommand{\gh}[1]{\bar{#1}}
\renewcommand{\aa}{\h a}
\newcommand{\ab}{\gh a}
\newcommand{\bb}{\gh b}
\newcommand{\cb}{\gh c}
\newcommand{\db}{\gh d}
\newcommand{\eb}{\gh e}
\newcommand{\fb}{\gh f}
\newcommand{\gb}{\gh g}
\newcommand{\pb}{\gh p}
\newcommand{\qb}{\gh q}

\begin{document}
\thispagestyle{empty}
\begin{flushright}
 \parbox{3.5cm}{KUCP-206\\ UT-02-22 \\
  {\tt hep-th/0204213}}
\end{flushright}

\vspace*{1cm}
\begin{center}
 {\Large
    Coset Construction of Noncompact $\Spin7$ and \G2 CFTs
 }
\end{center}

\vspace*{2cm} 
\begin{center}
 \noindent
 {\large Katsuyuki Sugiyama}

 \vspace{5mm}
 \noindent
 \hspace{0.7cm} \parbox{142mm}{\it
 Department of Fundamental Sciences, Faculty of Integrated Human Studies,
 Kyoto University, Yoshida-Nihonmatsu-cho,
 Sakyo-ku, Kyoto 606-8501, Japan.

 E-mail: {\tt sugiyama@phys.h.kyoto-u.ac.jp}
 }
\end{center}

 \vspace{5mm}
\begin{center}
 \noindent
 {\large Satoshi Yamaguchi}

 \vspace{5mm}
 \noindent
 \hspace{0.7cm} \parbox{142mm}{\it
 Department of Physics, Faculty of Science, University of Tokyo,
 Tokyo 113-0033, Japan.\\
 E-mail: {\tt yamaguch@hep-th.phys.s.u-tokyo.ac.jp}
}
\end{center}

\vspace{2cm}
\hfill{\bf Abstract\ \ }\hfill\ \\
We provide a new class of exactly solvable superconformal field theories 
that corresponds to type II compactification on manifolds with
exceptional holonomies. We combine $\Ncal=1$ Liouville field and $\Ncal=1$ 
coset models and construct modular invariant partition functions of
strings moving on these manifolds.
The resulting theories preserve spacetime
supersymmetry. Also we explicitly construct chiral currents 
in these models to realize consistent string theories.

\vspace{5mm}
\newpage

\section{Introduction}
Recently manifolds with exceptional holonomy are receiving much 
attention. These are $7$-dimensional manifolds with $G_2$ holonomy and 
$8$-dimensional manifolds with \Spin 7 holonomy. They provide 
interesting compactifications in string theories with minimal spacetime 
supersymmetry and are described by sigma models. To be consistent string 
vacua, the supersymmetric sigma models on these manifolds must be 
described by conformal field theories (CFTs). We can make use of 
worldsheet CFT techniques to describe the dynamics of these models. As 
is well-appreciated in the context of Calabi-Yau manifolds, conformal 
field theories and their chiral algebras are powerful tools to explore 
string theories. Exactly solvable models provide structural information 
and are certainly important starting points. Worldsheet description of 
string theories on these manifolds has been discussed and structures of 
their extended chiral algebras have been clarified.

Starting with the early work of Shatashvili and 
Vafa\cite{Shatashvili:1994zw}, there have been several papers about 
features of the superconformal field theories describing strings moving 
on manifolds with exceptional 
holonomies\cite{Figueroa-O'Farrill:1997hm,Gepner:2001px,Eguchi:2001xa,%
Sugiyama:2001qh,Blumenhagen:2001jb,Roiban:2001cp,Eguchi:2001ip,%
Blumenhagen:2001qx, Noyvert:2002mc,Roiban:2002iv,Naka:2002}.  They have derived 
the extensions of $\Ncal=1$ superconformal algebras by the tricritical 
Ising or Ising algebras from free field representations.  After their 
study of the algebras, it becomes a natural problem to look for explicit 
realizations.  Recent works on the subject include constructions of 
modular invariant partition functions for strings on (non)compact 
manifolds with special holonomies.  Explicit $G_2$ manifolds constructed 
so far are given by certain toroidal orbifolds.  Another large class 
of $G_2$ manifolds is supposed to result from anti-holomorphic 
$\Zb_2$ quotients of Calabi-Yau manifolds times a circle.  Also \Spin 7 
CFT has been proposed as the $\Zb_2$ involution of Calabi-Yau four-fold. 
These classes include models constructed by 
Joyce\cite{Joyce1,Joyce2,Joyce3}.

In this paper we present a new class of exactly solvable superconformal 
field theories which corresponds to certain point in the moduli space of 
type II compactification on exceptional holonomy manifolds. We construct 
modular invariant partition functions of strings compactified on 
noncompact manifolds with exceptional holonomies. We combine $\Ncal=1$ 
Liouville field and $\Ncal=1$ coset models so that resulting theories 
possess spacetime supersymmetry. Also 
we explicitly construct sets of extended chiral currents of these models 
to realize consistent string theories. The purpose of the paper is to 
propose new explicit examples of rational conformal field theories with 
exceptional holonomies.

\section{Modular invariant partition functions}
\subsection{Partition functions of \Spin7 models}
\label{spin7CFT}
In this subsection, we take, as an ansatz of a \Spin 7 conformal field
theory, a noncompact $c=12$ model described by a coset conformal field
theory
\begin{align}
 \Rb_{\phi}\times \frac{G\times \SO n}{H}.\label{G2internal}
\end{align}
$\Rb_{\phi}$ is the supersymmetric linear dilaton system
containing a free boson $\phi$ and a free fermion $\psil$. 
This part has $\Ncal=1$ superconformal symmetry and associated
 currents are expressed as the  Feigin-Fuchs representation
\begin{align}
 &T_{L}=-\frac12 (\del \phi)^2-\frac{Q}{2}\del^2\phi-\frac12 \psil\del\psil,
 \quad G_{L}=i\psil\del\phi+iQ\del\psil, \quad c_{L}=\frac32 + 3Q^2.
\label{2-2}
\end{align}
$Q$ is the background charge determined by the criticality
condition, that is, the total central charge should be $12$.

On the other hand, in the coset CFT $\Acal:=\frac{G\times \SO n}{H}$,
$G$ is a semi-simple Lie group and $H$ is a subgroup of $G$ 
 (for the detail of the coset CFT, see \cite{Halpern:1996js} and references there in).
When $n=\dim G-\dim H$ is satisfied, this coset CFT has $\Ncal=1$
superconformal symmetry\cite{Kazama:1989qp}. 
This condition leads to 
construct a consistent superstring model. In this subsection we will 
restrict ourselves to
the case $n=7$
and study the coset CFT $\Acal:=\frac{G\times \SO 7}{H}$.
We shall introduce currents of affine $G$ as $\Jh^{A}(z),\ A=1,\dots,\dim G$,
and describe the level $1$ affine \SO 7 algebra by six free fermions
 $\psi^{\ab}(z),\ \ab=1,\dots,7$ . Then currents  $\Jt^{\aa},\
 \aa=8,\dots,\dim G$
of affine $H$ are defined as 
the subalgebra of affine $G\times \SO 7$ in the form
\begin{align}
 \Jt^{\aa}=\Jh^{\aa}+\Jf^{\aa},\qquad
 \Jf^{\aa}:=-\frac{i}{2}f_{\aa\bb\cb}\psi^{\bb}\psi^{\cb}.\label{G2Jf}
\end{align}
$f_{ABC},\ A,B,C=1,\dots,\dim G $ is the structure constant of
$G$.  Throughout this subsection, we assume that $\Jh^{\aa}$'s, 
$\aa=8,\dots,\dim G$ are affine
$H$ generators.  By using these generators of affine algebras, 
currents of $\Ncal=1$ superconformal symmetry are
constructed by the standard method
\begin{align}
   T_{\Acal}
          &=\frac1k\left[
               \Jh^{\ab}\Jh^{\ab}-\frac{\kh}{2}\psi^{\ab}\del\psi^{\ab}
           +if_{\aa\bb\cb}\Jh^{\aa}\psi^{\bb}\psi^{\cb}
           -\frac12 f_{\ab\pb\qb}f_{\bb\pb\qb}\psi^{\ab}\del\psi^{\bb}
           -\frac14 f_{\ab\bb\pb}f_{\cb\db\pb}
                     \psi^{\ab}\psi^{\bb}\psi^{\cb}\psi^{\db}
                   \right],\nn\\
    G_{\Acal}
         &=\sqrt{\frac{2}{k}}\left[\psi^{\ab}\Jh^{\ab}
          -\frac i6f_{\ab\bb\cb}\psi^{\ab}\psi^{\bb}\psi^{\cb}
         \right],\qquad
   c_{\Acal}
        =\frac{3\kh}{2k}\dim(G/H)+\frac{1}{2k}f_{\ab\bb\cb}f_{\ab\bb\cb},
\label{2-4}
\end{align}
where $\kh$ is the level of affine $G$, $\k$ is supersymmetric level
defined by $\k:=\kh+\hc$, $\hc$ is the second Casimir of the adjoint
representation of $G$, defined by $f_{ACD}f_{BCD}=\hc\delta_{AB}$.

The character of the coset CFT can be obtained by branching relations.
Let $\Lambda$ be an integrable highest weight of affine $G$ and
$\chi^{G}_{\Lambda}(\tau)$ be the character of the representation with
highest weight $\Lambda$ . We express the representation of $\SO 7$ by
an index $s=0,1,2$ which labels basic, spinor and vector
representation, respectively. We denote the character of \SO 7 in the
representation $s$ as $\chi^{\SO 7}_{s}(\tau)$. As for the affine $H$,
we write the character of the representation with highest weight
$\lambda$ as $\chi^{H}_{\lambda}(\tau)$. With these notations, a module of
$\Acal$ is labelled by three indices $\Lambda,s,\lambda$ and the
character $\chi^{\Acal}_{\Lambda,s,\lambda}(\tau)$ of this module is 
obtained by the branching relation
\begin{align*}
 \chi^{G}_{\Lambda}(\tau)\chi^{\SO 7}_{s}(\tau)
   =\sum_{\lambda}\chi^{\Acal}_{\Lambda,s,\lambda}(\tau)
        \chi^{H}_{\lambda}(\tau).
\end{align*}

Now, we consider the type II string theory compactified by the CFT in
(\ref{G2internal}).  Because we are constructing the \Spin 7 CFT,
the resulting theory should be supersymmetric in spacetime, but this claim
is generally non-trivial.  When does this theory have spacetime
supersymmetry?  We investigate this problem in the following: Let us
denote the affine $H$ generated by the currents $\Jf^{\aa}$ in
(\ref{G2Jf}) by $\Hf$.  For the existence of spacetime supersymmetry the
size of this algebra $\Hf$ is crucial.  Generally, $\Hf$ is a subalgebra
of \SO 7, but it should be smaller to realize supersymmetry in
spacetime.  In fact we claim a proposition:
\begin{quote}
 If $\Hf \subset \G 2$ , the theory has spacetime supersymmetry.
\end{quote}
We will show this by demonstrating the partition function 
actually vanishes.

Let us first construct the partition function 
by the method of
\cite{Eguchi:2001ip}. When we take the light-cone gauge, 
the total theory
becomes product of two parts
$ \Rb_{\phi}\times \Acal.$
Next we introduce the building block $F^{\Spin7}_{\Lambda,\lambda}(\tau)$ as
a combination of characters
\begin{align*}
 F^{\Spin7}_{\Lambda,\lambda}(\tau)=
    \chi^{\Acal}_{\Lambda,2,\lambda}(\tau)\chi^{\isi}_{0}(\tau)
   +\chi^{\Acal}_{\Lambda,0,\lambda}(\tau)\chi^{\isi}_{1/2}(\tau)
   -\chi^{\Acal}_{\Lambda,1,\lambda}(\tau)\chi^{\isi}_{1/16}(\tau),
\end{align*}
where $\chi^{\isi}_{h}(\tau)$'s with 
$h=0,1/16,1/2$ are characters of Ising model
and represent contributions of $\psil$. With this block, 
we obtain the total partition function as
\begin{align*}
 Z^{\Spin7}(\tau,\taub)=\left(\sqrt{\im \tau}|\eta(\tau)|^2\right)^{-1}
\sum_{\Lambda,\lambda}\left|F^{\Spin7}_{\Lambda,\lambda}(\tau)\right|^2,
\end{align*}
where $\left(\sqrt{\im \tau}|\eta(\tau)|^2\right)^{-1}$ is the
contribution of $\phi$.
Also we can show the partition function 
$Z^{\Spin7}(\tau,\taub)$ 
vanishes.
This is equivalent to $F^{\Spin7}_{\Lambda,\lambda}(\tau)=0$. 
Under the assumption
$\Hf \subset \G2$, $F^{\Spin7}_{\Lambda,\lambda}(\tau)$
is rewritten as the form
\begin{align}
 F^{\Spin7}_{\Lambda,\lambda}&=\sum_{a=0,1}\xi^{G_2}_{a}(\tau)
\chi^{G\times G_2/H}_{\Lambda,a,\lambda}(\tau).\label{Spin7id}
\end{align}
Here $\chi^{G\times
G_2/H}_{\Lambda,a,\lambda}$'s are characters of the coset model
$G\times G_2/H$ defined by the branching relation
of affine \G2 characters $\chi^{G_2}_{a}$'s
\begin{align*}
 \chi^{G}_{\Lambda}(\tau)\chi^{G_2}_{a}(\tau)=\sum_{\lambda}\chi^{G\times
G_2/H}_{\Lambda,a,\lambda}(\tau)\chi^{H}_{\lambda}(\tau).
\end{align*}
On the other hand 
$\xi^{G_2}_{a}$'s $( a=0,1)$ are defined by characters 
$\chi^{\tri}_h(\tau)$ of 
tri-critical Ising model with $h=0,3/2,7/16,3/5,1/10,3/80$ 
\begin{align}
 &\xi^{G_2}_{0}(\tau):=\chi^{\isi}_{1/2}(\tau)\chi^{\tri}_{0}(\tau)
     +\chi^{\isi}_{0}(\tau)\chi^{\tri}_{3/2}(\tau)
    -\chi^{\isi}_{1/16}(\tau)\chi^{\tri}_{7/16}(\tau)\equiv 0,\nonumber\\
 &\xi^{G_2}_{1}(\tau):=\chi^{\isi}_{1/2}(\tau)\chi^{\tri}_{3/5}(\tau)
   +\chi^{\isi}_{0}(\tau)\chi^{\tri}_{1/10}(\tau)
 -\chi^{\isi}_{1/16}(\tau)\chi^{\tri}_{3/80}(\tau)\equiv 0. \label{xiG2}
\end{align}
Actually these vanish identically. 
It shows that $F^{\Spin7}_{\Lambda,\lambda}(\tau)\equiv 0$ and 
guarantees the spacetime supersymmetry.

We make a remark here: Eqs.(\ref{xiG2}) have typically appeared in the
partition functions of \G2
compactifications\cite{Eguchi:2001xa,Sugiyama:2001qh,Yamaguchi:2001kq}.
 Though we consider the \Spin7 compactifications, our models contain 
these factors.
This fact is related to singularities of our models and 
enhanced spacetime 
superconformal symmetry appears in their dual models as indicated in 
\cite{Yamaguchi:2001kq}.

A simple example of this type is realized 
with $G=\SO 7$ and $H=\G2$. The model
proposed in \cite{Eguchi:2001xa} is the special case of this example
with restriction $\kh=0$ .

\subsection{Partition functions of \G 2 models}
\label{G2CFT}
Let us turn to the \G2 compactifications using linear dilation system
and coset CFT. We take the light-cone gauge and consider the
transverse theory
\begin{align*}
 \Rb\times \Rb_{\phi}\times \frac{G\times \SO 6}{H},
\end{align*}
with $\dim G-\dim H=6$ . The character of 
$\Bcal:=\frac{G\times \SO 6}{H}$ is evaluated by the branching relation 
\begin{align*}
 \chi^{G}_{\Lambda}(\tau)\chi^{\SO6}_{s}(\tau)
   =\sum_{\lambda}\chi^{\Bcal}_{\Lambda,s,\lambda}(\tau)
             \chi^{H}_{\lambda}(\tau),
\end{align*}
where the index $s=0,1,2,3$ represents the affine \SO6 representation and
each $\chi^{\SO6}_{s}$ is the character of affine \SO6. Other notations
are the same as the \Spin7 case. We introduce a  set of building blocks
$F^{G_2}_{\Lambda,\lambda}$ as a combination of characters
\begin{align*}
 F^{G_2}_{\Lambda,\lambda}(\tau)=
    \chi^{\Bcal}_{\Lambda,2,\lambda}(\tau)\chi^{\SO2}_{0}(\tau)
   +\chi^{\Bcal}_{\Lambda,0,\lambda}(\tau)\chi^{\SO2}_{2}(\tau)
   -\chi^{\Bcal}_{\Lambda,1,\lambda}(\tau)\chi^{\SO2}_{1}(\tau)
   -\chi^{\Bcal}_{\Lambda,3,\lambda}(\tau)\chi^{\SO2}_{3}(\tau).
\end{align*}
$\chi^{\SO2}_{s}$ is the character of affine $\SO2$ and has
contributions 
of $\psil$ and the free fermion in the transverse  direction 
in spacetime.
By using this building block, 
we can obtain the partition function as
\begin{align*}
 Z^{G_2}(\tau ,\bar{\tau})=(\sqrt{\im\tau}|\eta(\tau)|^2)^{-2}
  \sum_{\Lambda,\lambda}|F^{G_2}_{\Lambda,\lambda}(\tau)|^2.
\end{align*}
The factor $(\sqrt{\im\tau}|\eta(\tau)|^2)^{-2}$ represents 
contributions of $\phi$ and a spacetime  boson in $1$ dimensional 
transverse direction.

As for the spacetime supersymmetry in this compactification, we claim
the proposition:
\begin{quote}
 If $\Hf\subset \SU 3$, this theory has supersymmetry in spacetime.
\end{quote}
Actually, when $\Hf\subset \SU 3$ the block 
$F^{G_2}_{\Lambda,\lambda}(\tau)$
can be rewritten by using the characters  
$\chi^{G\times \SU3/H}_{\Lambda,a,\lambda}$'s of
the coset $G\times \SU3/H$,
\begin{align*}
 F^{G_2}_{\Lambda,\lambda}(\tau)=\sum_{a=-1,0,1}\xi^{\SU3}_{a}(\tau)
                  \chi^{G\times \SU3/H}_{\Lambda,a,\lambda}(\tau).
\end{align*}
$\xi^{\SU3}_{a}$'s are functions constructed from 
\SU2 classical theta functions  $\Theta_{m,k}$'s
\begin{align}
 \xi^{\SU3}_{a}(\tau)
   =\frac{1}{\eta(\tau)^2}\sum_{s\in\Zb_4}(-1)^{s}\Theta_{6+4a-3s,6}(\tau)
     \Theta_{s,2}(\tau) \equiv 0.\label{su(3)holonomyidentity}
\end{align}
We can show this set of functions vanishes identically.
The most simple example is illustrated with
$G=\SU2\times\SU2\times\SU2$ and $H=\SU2$. A series of models in
\cite{Eguchi:2001xa} is included in this example.
Another typical example is realized with $G=G_2$ and H=SU(3).

\section{Currents of the extended algebras}
\subsection{Currents of \Spin7 CFT algebra}
In this subsection, we will study currents of extended 
superconformal algebra associated with $8$ dimensional \Spin 7 manifold.
The extended symmetry algebra has been found in paper
\cite{Shatashvili:1994zw}.
In addition to a set of $\Ncal=1$ superconformal currents $(T,G)$, 
it contains operators $(\tilde{X},\tilde{M})$ with spins $(2,3/2)$.
$\tilde{X}$ is the energy momentum tensor 
for the $c=1/2$ model. 

We construct these currents in our coset models 
discussed 
in section \ref{spin7CFT}. 
The super-stress tensor $T$ and $G$ are given
as sums of $\Ncal=1$ Liouville parts $T_L$, $G_L$ and coset parts 
$T_{\cal A}$, $G_{\cal A}$ associated to
${\cal A}=\frac{G\times SO(7)}{H}$ with $\dim G-\dim H=7$
\begin{align*}
 T=T_{L}+T_{\Acal},\qquad G=G_{L}+G_{\Acal}.
\end{align*}
Their explicit formulae are expressed in Eqs.(\ref{2-2})(\ref{2-4}).
Next we propose that the current $\Xt$ is simply given 
by combining fermionic fields of Liouville and coset theories
\begin{align}
 \Xt=-\frac16 \Phi_{\ab\bb\cb}\psi^{\ab}\psi^{\bb}\psi^{\cb}\psil
+\frac{1}{24}\Phis_{\ab\bb\cb\db}\psi^{\ab}\psi^{\bb}\psi^{\cb}\psi^{\db}
-\frac12 \psi^{\ab}\del \psi^{\ab}-\frac12 \psil\del\psil,\qquad
\ab,\bb,\dots=1,\dots,7.\label{xt}
\end{align}
In this equation, we use the structure constants of the octonion 
$\Phi_{\ab\bb\cb}$ and its hodge dual $\Phis_{\ab\bb\cb\db}:=\frac16 
\epsilon_{\ab\bb\cb\db\eb\fb\gb}\Phi_{\eb\fb\gb}$. The form of $\Xt$ in 
Eq.(\ref{xt}) satisfies the following consistency conditions:
\begin{itemize}
 \item When $\Hf\subset 
G_2$, $\Xt$ is actually an operator in the coset theory $\Acal$, that is, 
the condition $\Jt^{\aa}(w)\Xt(z)\sim \text{(regular)}$ is satisfied.
 \item $\Xt$ itself has the OPE of the stress tensor in the 
 critical Ising model. The appearance of this statistical model is
       essential in the reduction of holonomy from $SO(8)$ to 
\Spin 7 through a relation $SO(8)$/\Spin 7 $\cong$ (Ising model).
\end{itemize}

The remaining current $\Mt$ is obtained by the OPE 
$G(z)\Xt(w)\sim \frac{G(w)/2}{(z-w)^2}+\frac{\tilde{M}(w)}{z-w}$ as
\begin{align}
 \Mt=\frac12\sqrt{\frac2k}\Bigg[&
-\Phi_{\ab\bb\cb}\Jh^{\ab}\psi^{\bb}\psi^{\cb}\psil
+\frac13 \Phis_{\ab\bb\cb\db}\Jh^{\ab}\psi^{\bb}\psi^{\cb}\psi^{\db}
-\Jh^{\ab}\del\psi^{\ab}
+\del\Jh^{\ab}\psi^{\ab}\nn\\
&+\frac{i}{2}f_{\pb\ab\bb}\Phi_{\pb\cb\db}
               \psi^{\ab}\psi^{\bb}\psi^{\cb}\psi^{\db}\psil
-if_{\pb\qb\ab}\Phi_{\pb\qb\bb}\del\psi^{\ab}\psi^{\bb}\psil
-\frac{i}{6}f_{\pb\ab\bb}\Phis_{\pb\cb\db\eb}
               \psi^{\ab}\psi^{\bb}\psi^{\cb}\psi^{\db}\psi^{\eb}\nn\\
&+\frac{i}{2}\left(f_{\pb\qb\ab}\Phis_{\pb\qb\bb\cb}-f_{\ab\bb\cb}\right)
          \del\psi^{\ab}\psi^{\bb}\psi^{\cb}
\Bigg]
+\frac16 \Phi_{\ab\bb\cb}\psi^{\ab}\psi^{\bb}\psi^{\cb}i\del\phi
-\frac12 \del \psil i\del\phi
+\frac12 \psil i\del^2 \phi.
\end{align}
We have checked that these currents $(T,G,\Xt,\Mt)$ actually satisfy the 
\Spin 7 CFT algebra for the example 
with $G=\SO7$ and $ H=G_2$. For general coset cases with 
$\Hf\subset G_2$, we propose
that the set of these currents 
also satisfies the \Spin 7 CFT algebra as a conjecture.

\subsection{Currents of \G2 CFT algebra}

We take a seven manifold with a $G_2$ holonomy. The $G_2$ 
structure on this manifold is given by a closed $G_2$ 
invariant $3$-form $\Phi$. 
By including this operator, an extended algebra has been 
constructed in the paper \cite{Shatashvili:1994zw}.
In addition to a set of stress tensor $T$ and its superpartner 
$G$, 
the algebra contains sets of chiral currents $(K,\Phi)$ 
with spins $(2,3/2)$ and $(X,M)$ with spins $(2,5/2)$. 
The $X$ is related with a dual $4$-form $\ast\Phi$ and $(X,\Phi) $
is a set of currents of $\Ncal=1$ additional superconformal 
algebra. 
The extra conformal algebra is isomorphic to 
the algebra of the tricritical Ising model with 
central charge $7/10$ and 
plays an essential role in 
the reduction of holonomy of the manifold from $SO(7)$ to $G_2$ 
through a relation $SO(7)/G_2\cong$ (Tricritical Ising).

In this subsection, we construct these currents in the coset $G_2$ model 
discussed in section 2.2. First the set of currents $(T,G)$ is 
constructed as combinations of Liouville parts $T_L$, $G_L$ 
and coset parts $T_{\cal B}$, $G_{\cal B}$ 
associated with 
${\cal B}=\frac{G\times SO(6)}{H}$ ($\dim G-\dim H =6$)
\begin{align*}
 &T=T_{L}+T_{\Bcal},\qquad G=G_{L}+G_{\Bcal}.
\end{align*}
Next we take the current $\Phi$ as 
a combination of a Liouville fermion and 
fermionic fields of the coset model
\begin{align*}
 &\Phi=\frac16 A_{\ab\bb\cb}\psi^{\ab}\psi^{\bb}\psi^{\cb}
         +\frac12 H_{\ab\bb}\psi^{\ab}\psi^{\bb}\psil .
\end{align*}
Other currents $X$, $K$, $M$ are constructed
 by the OPEs $\Phi(z)\Phi(w)\sim -\frac{7}{(z-w)^3}+\frac{6}{z-w}X(w)$, 
$G(z)\Phi(w)\sim \frac{1}{z-w}K(w)$, $G(z)X(w)\sim -\frac{1}{2}
\frac{1}{(z-w)^2}G(w)+\frac{1}{z-w}M(w)$
respectively  
\begin{align*}
 &X=-\frac{1}{24}\Hs_{\ab\bb\cb\db}\psi^{\ab}\psi^{\bb}\psi^{\cb}\psi^{\db}
   +\frac{1}{6}\As_{\ab\bb\cb}\psi^{\ab}\psi^{\bb}\psi^{\cb}\psil
 +\frac12 \psi^{\ab}\del\psi^{\ab}+\frac12 \psil\del\psil ,\\
 &K=\frac12\sqrt{\frac 2k}\Bigg[
    A_{\ab\bb\cb}\Jh^{\ab}\psi^{\bb}\psi^{\cb}
    +2 H_{\ab\bb}\Jh^{\ab}\psi^{\bb}\psil
    -\frac{i}{2}f_{\pb\ab\bb}A_{\pb\cb\db}\psi^{\ab}\psi^{\bb}\psi^{\cb}\psi^{\db}
  \nn \\& \hspace{2cm}
    +if_{\pb\qb\ab}A_{\pb\qb\bb}\del\psi^{\ab}\psi^{\bb}
    -if_{\pb\ab\bb}H_{\pb\cb}\psi^{\ab}\psi^{\bb}\psi^{\cb}\psil
    +if_{\pb\qb\ab}H_{\pb\qb}\del\psi^{\ab}\psi_{\phi}
    \Bigg]+\frac12 H_{\ab\bb}\psi^{\ab}\psi^{\bb}i\del\phi,\\
 &M=\frac12 \sqrt{\frac 2k}\Bigg[
     \As_{\ab\bb\cb}\Jh^{\ab}\psi^{\bb}\psi^{\cb}\psil
     -\frac13\Hs_{\ab\bb\cb\db}\Jh^{\ab}\psi^{\bb}\psi^{\cb}\psi^{\db}
     +\Jh^{\ab}\del\psi^{\ab}
     -\del\Jh^{\ab}\psi^{\ab}
   \nn\\&\hspace{2cm}
     -\frac{i}{2}f_{\pb\ab\bb}\As_{\pb\cb\db}
           \psi^{\ab}\psi^{\bb}\psi^{\cb}\psi^{\db}\psil 
     +if_{\pb\qb\ab}\As_{\pb\qb\bb}\del\psi^{\ab}\psi^{\bb}\psil
     +\frac{i}{6}f_{\pb\ab\bb}\Hs_{\pb\cb\db\eb}
           \psi^{\ab}\psi^{\bb}\psi^{\cb}\psi^{\db}\psi^{\eb}
   \nn\\&\hspace{2cm}
     +\frac{i}{2}\left(-f_{\pb\qb\ab}\Hs_{\pb\qb\bb\cb}+f_{\ab\bb\cb}\right)
                  \del\psi^{\ab}\psi^{\bb}\psi^{\cb}
    \Bigg]+\frac12 \del\psil i\del\phi -\frac12 \psil i\del^2\phi
      -\frac16\As_{\ab\bb\cb}\psi^{\ab}\psi^{\bb}\psi^{\cb}i\del\phi,\\
 &\ab,\bb,\dots=1,\dots,6.
\end{align*}
Here we used notations
\begin{align*}
 &A_{\ab\bb\cb}:=\Phi_{\ab\bb\cb},\qquad H_{\ab\bb}:=\Phi_{\ab\bb 7},\qquad
   \As_{\ab\bb\cb}:=\frac{1}{6}
              \epsilon_{\ab\bb\cb\db\eb\fb}A_{\db\eb\fb},\qquad
   \Hs_{\ab\bb\cb\db}:=\frac{1}{2}\epsilon_{\ab\bb\cb\db\eb\fb}H_{\eb\fb}.
\end{align*}
For general coset cases, 
we propose that these currents satisfy the full $G_2$ CFT algebra 
as a conjecture when $\Hf \subset \SU3$ is satisfied.

\section{Conclusion}\label{conclusion}
In this paper, we construct the \Spin 7 and $G_2$ CFTs combining
 $\Ncal=1$ Liouville and supersymmetric coset models. 
It provides a new class of exactly solvable superconformal field
theories that corresponds to type II strings compactified on 
exceptional holonomy manifolds. 

We construct 
modular invariant
partition functions. 
It is shown that they vanish and 
we make sure that the resulting theories 
possess spacetime supersymmetry.
These noncompact models include 
factors $\xi^{G_2}_a$, $\xi^{SU(3)}_a$ 
in partition functions as if 
the models have twice as mamy supersymmetry as expected.
When the target manifolds become singular, 
their dual field theories become interacting superconformal 
field theories.
It suggests the ``holographic dual'' theories
of these string models
are superconformal as indicated in \cite{Yamaguchi:2001kq} 
and extra supercharges correspond to superconformal $S$ 
generators in the dual theories.
It is interesting to apply present approach to investigation of 
properties in dual field theories. 

We also explicitly construct 
the sets of the \Spin 7 and $G_2$ CFT currents in our models 
to realize consistent string theories. 
Among the class of the models considered in subsections 2.1 and 3.1,
$\Rb_{\phi}\times (SO(7) \times SO(7)) /G_2 $ is the most typical example. 
For $G_2$ holonomy case, 
$\Rb_{\phi}\times   (G_2 \times SO(6)) /SU(3) $ 
is possible in our coset construction. 
We will make a remark here: In our paper we mainly discuss 
exceptional holonomy cases and investigate their coset 
construction in noncompact models. 
But our construction is not restricted 
to these exceptional holonomy manifolds. It is 
also applicable to other special holonomy cases. 
For example, we can propose 
$\Rb_{\phi}\times (SU(3) \times SO(5)) /SU(2) $,  
$\Rb_{\phi}\times (SU(2) \times SO(3)) $
respectively
as SU(3), SU(2) holonomy models.
By considering a string of reduction of 
holonomies, we can show a cascade of special holonomy groups. 
Typically it appears in these four cases. 
Their fermionic parts are described by $SO$ groups. 
When the holonomies are reduced, 
the dimensions of manifolds decrease and  
the $SO$ group parts are changed gradually as 
$SO(7)\rightarrow SO(6)\rightarrow SO(5)\rightarrow SO(3)$. 
On the other hand, bosonic parts of these cosets are transformed as 
$SO(7)/G_2\rightarrow G_2/SU(3)\rightarrow 
SU(3)/SU(2)\rightarrow SU(2)$. 
It represents a string of statistical models 
indicated in \cite{Sugiyama:2001qh}, 
$SO(7)/G_2\cong$(Tricritical Ising), 
$G_2/SU(3)\cong$(3-state Potts), $SU(3)/SU(2)\cong U(1)$ 
and their central charges increase. 
In these models holonomies are actually reduced as 
\Spin 7 $\rightarrow G_2\rightarrow SU(3)\rightarrow SU(2)$. 
It is an interesting problem to study these strings based on 
sigma models or 
gauged WZW models that have an explicit picture of target space 
geometries. 

\subsection*{Acknowledgement}
We would like to thank Tohru Eguchi, Yuji Sugawara and Christian 
R\"omelsberger for useful discussions and comments.

The work of K.S. is supported in part by the Grant-in-Aid from the Ministry 
of Education, Science, Sports and Culture of Japan
(\#14740115).
The work of S.Y. is supported in part by Soryushi Shogakukai.
\providecommand{\href}[2]{#2}\begingroup\raggedright\endgroup
\end{document}